\documentclass[a4paper,notitlepage,11pt]{article}%
\usepackage{amsmath}
\usepackage{amsfonts}
\usepackage{amssymb}
\usepackage{graphicx}%
\providecommand{\U}[1]{\protect\rule{.1in}{.1in}}

\begin{document}

\title{Can intrinsic noise induce various resonant peaks?}
\author{J.J. Torres, J. Marro, and J.F. Mejias$^{\ast}$\\Institute \textit{Carlos I} for Theoretical and Computational Physics,\\and \textit{Departamento de Electromagnetismo y F\'{\i}sica de la Materia},\\University of Granada, Spain. $^{\ast}$Also at Centre for Neural\\Dynamics, University of Ottawa, Canada.}
\maketitle

\begin{abstract}
We theoretically describe how weak signals may be efficiently transmitted
throughout more than one frequency range in noisy excitable media by kind of
\textit{stochastic multiresonance}. This serves us here to reinterpret recent
experiments in neuroscience, and to suggest that many other systems in nature
might be able to exhibit several resonances. In fact, the observed behavior
happens in our (network) model as a result of competition between (1) changes
in the transmitted signals as if the units were varying their activation
threshold, and (2) adaptive noise realized in the model as rapid
activity--dependent fluctuations of the connection intensities. These two
conditions are indeed known to characterize heterogeneously networked systems
of excitable units, e.g., sets of neurons and synapses in the brain. Our
results may find application also in the design of detector devices.\medskip

\noindent\noindent\underline{PACS numbers}: 05.40.-a; 43.50.+y;
87.10.Mn;87.19.ln; 87.19.lt

\noindent\underline{Key words}: excitable media; undamped signal transmission;
stochastic multiresonance; connection fatigue; short--time plasticity;
multiplicative noise; networks.

\end{abstract}

Some systems in nature are known to process efficiently weak signals in noisy
environments. A novel mechanism that explains such ability is known as
\textit{stochastic resonance} (SR). This is associated with the occurrence of
a peak or bell--shaped dependence in the transfer of information through an
excitable system as a noise source is conveniently tuned. More specifically,
low or slow noise impedes detecting a relatively weak signal but, as the noise
raises, the system eventually responds correlated with the signal, which shows
as a peak of information transfer. The signal is again obscured at higher
noise levels. This has been reported to occur in different settings, including
electronic circuits, ring lasers, crayfish mechanoreceptors, ion channels,
sensory neurons, hippocampus, brain stem, and cortical areas
\cite{WMnature95,hanggi09,LindnerGO04,wel04}. An intriguing issue raised is
whether a given system may filter with gain in different noise regimes, which
would have technological application. After the first proposal of stochastic
\textit{multiresonance} (SMR) \cite{rubi97}, the existence of two or more
resonant peaks has been predicted for single--mode lasers \cite{gui03},
surface phenomena \cite{zhao05}, biological diversity \cite{raul06} and
intracellular calcium oscillations in hepatocytes \cite{zhang04,zhang08}, and
it has also been described in somewhat more abstract settings
\cite{kim01,maty03,luo03,volkov05,hong05,barbi05}. Though there is no definite
claim for experimental evidence of SMR yet, two recent sets of experimental
data \cite{tsin00,yasuda08} admit such interpretation.

Here we demonstrate that a single resonant mechanism may indeed help in
transmitting signals throughout different noise frequencies. More
specifically, we use an explicit mathematical model ---based on independent
familiar empirical descriptions for both \textit{neuron} units and their
\textit{synaptic} links--- to reveal the existence of a double resonance in an
experiment concerning the human tactile blink reflex \cite{yasuda08}. Our
model behavior is also consistent with recent reports on the transfer of
information with different frequencies in the hippocampus \cite{colgin09}. On
the other hand, the model here allows one to modify the separation between the
two peaks in one order of magnitude or more, and it may admit generalization
to show more than two resonances, which makes the \textquotedblleft
device\textquotedblright\ very versatile.

Our main result suggests looking for SMR in nature as part of a needed effort
to better understand how the details in excitable systems influence
transmission. Previous studies of SR and SMR in nonlinear settings most often
involved a source of controlled,
additive noise getting rid of correlations. The case of an\textit{ intrinsic},
therefore uncontrolled, noise resulting from inherent activity in the medium
is even more interesting and likely to occur in nature. In a cortical region,
for instance, a given neuron may receive, in addition to the (weak) signal of
interest, uncorrelated, i.e., just noisy signals from other neurons at
frequencies that vary in time during the normal operation of the system.
Following previous efforts \cite{antes,mt}, we consequently investigated the
possibility of having SMR associated with the fact that both the main signal
and the noise transmit through dynamic connections, e.g., synapses, whose
weights change with time and, therefore, constantly modulate transmission. We
found in this way that short--term activity--dependent \textquotedblleft
fatigue plasticity\textquotedblright\ ---such as synaptic \textit{depression}
and \textit{facilitation }which is known to modify the neural response causing
complex behavior \cite{tsodyks97,abbott97,buia05,torresnc08,mejias08}--- may
indeed produce SMR in a model of neural media in agreement with recent
observations. The setting in this paper, which may be seen as an application
of a general study within a biological context \cite{mt}, intends both to
serve as a simple illustration of our point and to make contact with a
specific experiment. However, it is sensible to anticipate that the main
outcome here may hold rather generally in excitable systems, given that these
seem to share all the relevant features in our model \cite{NOSexcit2,NOSexcit}.

Consider a networked system in which units, say neurons, receive: (\textit{i})
a weak signal from other \textit{brain} areas and/or from the \textit{senses}
or whatever external terminals and, in addition to this, (\textit{ii})
uncorrelated, noisy signals from other units. The latter signals will be
portrayed by means of action potentials (AP) ---from the many\textit{
presynaptic} neurons to the \textit{postsynaptic} neuron--- whose rates follow
a Poisson distribution with mean $f$ \cite{note0}. Besides the stochasticity
this implies, we shall imagine the neurons connected by dynamic,
activity--dependent links. To be specific, we shall adopt the model of
\textit{dynamic synapses} in \cite{N1}. That is, any synaptic link, say $i,$
is essentially stochastic which is implemented assuming it composed of an
arbitrary number, $M_{i},$ of functional contacts, each releasing its
transmitter content with probability $u$ when an AP signal from other units
arrives. Furthermore, to implement excitability (and, more important here,
kind of threshold fickleness), the contact is assumed to become inactive
afterwards for a time interval, $\tau;$ this is a random variable with
exponential distribution $p_{t}\left(  \tau\right)  $ of mean $\tau
_{\text{rec}}$ at time $t.$ Therefore, each activation event, i.e., the
arrival of an AP at $i$ at time $t_{i}$ generates a (postsynaptic) signal,
$I_{i}(t)$, which evolves according to%
\begin{equation}
\frac{dI_{i}(t)}{dt}=-\frac{I_{i}(t)}{\tau_{\text{in}}}+\sum_{\ell=1}^{M_{i}%
}J_{i,\ell}\,x_{i,\ell}\,\delta(t-t_{i}). \label{eq1}%
\end{equation}
Here, $J_{i,\ell}$ is the modification in the signal produced by the AP in
contact $\ell$ after the release event, and $x_{i,\ell}\left(  t\right)  =1$
when the contact is activated, which occurs with probability $u\left[
1-p_{t}\left(  \tau\right)  \right]  ,$ and $0$ otherwise. The time constant
$\tau_{\text{in}}$ is a measure of the transmission duration (of order of
milliseconds for a known type of fast postsynaptic receptors). For $N$ units,
the total postsynaptic signal is $I_{N}(t)=\sum_{i=1}^{N}I_{i}(t)$. We also
assume, as in \cite{N1}, that both the number and the strength of functional
contacts that a presynaptic unit $i$ establishes, namely, $M_{i}$ and
$J_{i,\ell}$ vary with $i$ according to Gaussian distributions of mean and
standard deviation $\left(  M,\Delta_{M}\right)  $ and $\left(  J,\Delta
_{J}\right)  ,$ respectively. To compare with a specific experiment, we assume
$M=50\pm0.1$ contacts, $J=0.3\pm0.1$ mV, $u=0.5$, $\tau_{in}=1$ ms and
$\tau_{\text{rec}}=500$ ms \cite{nota29bis}. Just for simplicity \cite{nota1},
we consider a weak, low--frequency sinusoidal signal $S(t)\equiv d_{s}%
\cos(2\pi f_{s}t)$ which is transmitted to the (postsynaptic) unit to monitor
the corresponding response and the conditions in which resonance occurs. With
this aim, we then compute the generated voltage, $V(t),$ assuming a generic
dynamics of the form:%
\begin{equation}
\frac{dV}{dt}=F\left(  V,I_{N},S\right)  , \label{dynv}%
\end{equation}
where the function $F$ is to be determined.

Once the links are determined, specifying $F$ means adopting a model for each
unit. A familiar choice is the \textit{integrate--and--fire }(IF) model in
which $F$ is linear with $V$ \cite{NR1}. This assumption of a fixed firing
threshold is a poor description for most purposes \cite{NR2}, however.
Instead, one could assume a networked stochastic set of (postsynaptic) units
---e.g., a convenient adaptation of the network model in
\cite{NOSexcit2,NOSexcit}--- but, for the shake of simplicity, we shall
inspire ourselves here in the FitzHugh--Nagumo (FHN) model \cite{NR3}. The
excitability is then implemented assuming that the thresholds for neuron
shoots constantly adapt to the input current, which is realistic for neuron
media \cite{N2}. Summing up, the unit dynamics is
\begin{equation}
{\frac{dV(t)}{{\epsilon\,}dt}}{=V(t)\left[  V(t)-a\right]  \left[
1-V(t)\right]  -W(t)}+\frac{\tilde{S}(t)}{{\epsilon\,\tau_{m}}}, \label{S1}%
\end{equation}
where $\tilde{S}(t)=S(t)+\rho I_{N}(t)$ is the input, with $\rho$ a resistance
that transforms the current $I_{N}$ into a voltage. $W(t),$ which stands for a
(slow recovery) variable accounting for the refractory time of the unit,
satisfies:%
\begin{equation}
{\frac{dW(t)}{dt}}{=b\,V(t)-c\,W(t).} \label{S2}%
\end{equation}
In order to compare with the experiment of interest, we shall take $a=0.001,$
$b=3.5\,ms^{-1},$ $c=1\,ms^{-1},$ and $\epsilon=1000\,ms^{-1}$ which makes the
(dimensionless) voltage $V(t)=1$ to correspond to $100\,mV$ and the time
variable to be within the $ms$ range. We further assume a membrane resistance
$\rho=0.1\,G\Omega/mV$ and a time constant $\tau_{m}=10~ms$ both within the
physiological range \cite{NR4}.

\begin{figure}
[tbh]
\begin{center}
\includegraphics[
height=7.9166cm,
width=11.071cm
]%
{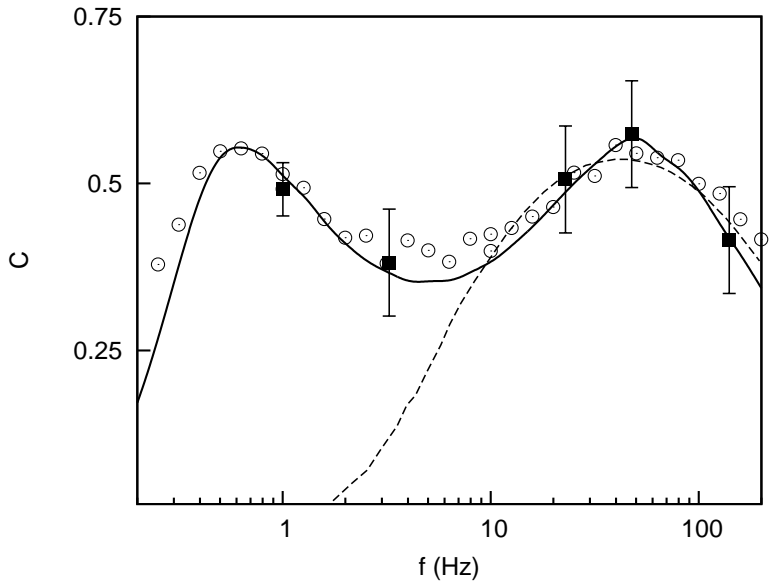}%
\caption{The function $C$ (as defined in the text but scaled, arbitrary units
so that it measures relative variations of the relevant correlation),
according to the experimental data in \cite{yasuda08} (full squares with their
error bars) and our prediction (solid line). The empty symbols are the
response when the signal $S\left(  t\right)  ,$ instead of the sinusoidal
(therefore, time correlated) one producing the solid line, consists of a train
of (uncorrelated) Poissonian pulses which, as the main signal, also endure the
model synaptic dynamics; the only noticeable change is that the response
results more noisy in this case due to extra randomness. The dashed line
corresponds to the interpretation of these data given in \cite{yasuda08}. (The
parameter values used in these plots are well within the corresponding
physiological range; see \cite{nota2} and the main text for details.) }%
\label{figure1}%
\end{center}
\end{figure}

The degree of correlation between the input signal and the output $V(t)$ is
defined as
\begin{equation}
C=\left\langle S(t)\nu(t)\right\rangle \equiv\frac{1}{T}\int_{t_{0}}^{t_{0}%
+T}S(t)\nu(t)dt. \label{eq2}%
\end{equation}
Here, $\nu(t)$ is the instantaneous firing rate of the postsynaptic unit, that
is, the average number of AP's generated at time $t$ as a consequence of input
$\tilde{S}.$ (In practice, the average is over a set of different postsynaptic
AP's trains generated under the same experimental conditions.) The function
$C\left(  f\right)  $ that follows from this is shown as a solid line in
figure \ref{figure1}.%

As said above, previous studies illustrated SR as a peak of $C$ when one
varies the level of a noise which is apropos injected in the system. In our
excitable model system, however, is the synaptic current $I_{N}(t)$ ---and not
an external noise--- what directly affects dynamics. Tuning the level of noise
now means increasing the frequency $f$ of the uncorrelated AP's that are
responsible for the generation of $I_{N}(t).$ The noise embedded in the AP's
trains does not directly affect the unit, and this has a strong consequence on
the shape of $C.$ That is, SMR is then a consequence of interplay between
short--term (\textit{synaptic}) plasticity and threshold variability
associated to dynamics (\ref{dynv}).

To be more specific, let us write the total signal as $I_{N}=\bar{I}_{N}%
\pm\sigma_{I}$, where $\bar{I}_{N}>0.$ In the IF model (fixed threshold),
$\bar{I}_{N}$ tends to reduce the voltage needed for the generation of an AP,
so that the excitability of the neuron increases with $\bar{I}_{N}.$ In the
FHN model, however, the main effect of $\bar{I}_{N}$ is to move the stationary
solution of system (\ref{S1})--(\ref{S2}) (and the $V$ nullcline) towards more
positive values, so that both the resting voltage value and the voltage
threshold become more positive. Then, for the range of $\bar{I}_{N}$ values of
interest here, the neuron excitability depends more on the fluctuation
$\sigma_{I}$ than on $\bar{I}_{N}.$ On the other hand, for dynamic synapses,
$\sigma_{I}=\sigma_{I}(f)$ has a non-monotonic dependence on $f$ ---it first
increases from 0, reaches a maximum, say $f^{\ast},$ and then decreases with
increasing $f$ to cero again. As a result, if the level of fluctuations at
$f^{\ast}$ is such that the unit is above threshold, there will be two
frequency values for which ---according to familiar arguments
{\cite{WMnature95}---} fluctuations may eventually overcome the potential
barrier between silent and firing states, which results in two resonant peaks.

The model here allows one to understand, even semi--quantitatively recent data
by Yasuda \textit{et al.} \cite{yasuda08} showing how short--term synaptic
depression causes stochastic resonance in human tactile blink reflex. These
authors monitored an input--output correlation function between tactile signal
and blink reflexes associated with postsynaptic responses of neurons in the
caudal pontine reticular nucleus (PnC) of the brainstem \cite{friauf94}. In
addition to the (weak) tactile signal, these neurons received uncorrelated
auditory inputs that are viewed as a noise background. Yasuda \textit{et al.}
then concluded that, for realistic short--term depression parameters, the
postsynaptic neuron acts as an error--free detector. That is, the value of the
input--output correlation function is maintained at optimal high values for a
wide range of background noisy rates.

A close inspection of the Yasuda \textit{et al.} data from the perspective
above reveals some significant discrepancies of the fit in that study at low
noise rates. That is, while experiments for low noise rates show a high
input--output correlation level (see Table I in Ref.\cite{yasuda08}), the
theory ---based on the oversimplified, linear IF neuron model with fixed
firing threshold--- they use to interpret and compare with their data does not
predict SMR but a very low level of correlation at low frequency which is not
supported by data (consequently, the authors in \cite{yasuda08} excluded their
low frequency data from their analysis). This is shown in figure
\ref{figure1}. The disagreement may be justified by accepting that, at such
low rate and due to the high neuron threshold in PnC area, the auditory noise
is not enough to evoke a postsynaptic response correlated to the signal. So
the high level of the correlation observed can only be understood by the
effect of noise coming from other brain areas. Those authors did not study
this additional noise source, however, so that the question of whether other
brain areas play a role here remains unanswered. On the other hand, if such a
noise is relevant, its effect should be a constant noise added to the auditory
noise. Therefore, it should induce a constant increment in the noise level,
which cannot explain two local maxima apparently observed in the experimental
correlation data (fig.\ref{figure1}) at noise levels around $1$ and $50$ Hz,
respectively.

The drawing of data in the plot of figure \ref{figure1}, particularly those
for small $f$ undervalued in \cite{yasuda08}, requires a comment \cite{nota2}.
That is, one needs to use a specific relationship between the auditory noise
and $f.$ Let us assume \cite{dayan} that the firing rate of neuron $i$ in the
PnC area induced by an auditory input $A$ is $f_{i}=f_{0}+\alpha
A\Theta(A-A_{i}),$ where $f_{0}$ is the level of activity in absence of any
input, $\alpha$ is a constant, $\Theta(x)$ is the step function, and $A_{i}$
is the minimum input needed to induce non-spontaneous activity in neuron $i$.
The known variability of the firing thresholds in most, e.g., PnC neurons
\cite{friauf94}, suggests one to sample $A_{i}$ from a Gaussian distribution
with mean $A_{0}$ and variance $\sigma_{A}^{2}.$ It then follows that the mean
firing rate (in Hz) induced in the PnC area by an auditory input $A$ (in dB)
is
\begin{equation}
f=f_{0}+\frac{\alpha A}{2}\left\{  1+\mbox{erf}\left(  \frac{A-A_{0}}{\sqrt
{2}\sigma_{A}}\right)  \right\}  . \label{nlr}%
\end{equation}
This, which generalizes the linear relationship used in \cite{yasuda08} within
a restricted range, transforms all levels of auditory noise (between 30 and 90
dB in the experiment of interest) into the frequency domain. For $A\gg A_{0},$
(\ref{nlr}) reduces to a linear relation.

Summing up, our model system predicts two maxima, and not only one, in the
transfer of information during the specific situation that we show in
fig.\ref{figure1}. This, to be interpreted as SMR phenomena, provides
\textit{a priori} a good description of the only sufficiently detailed
observation we know, namely, it fits \underline{all} the data in
\cite{yasuda08}, and it is also in qualitative agreement with several
predictions, as mentioned above, and with the global behavior reported in
various experiments \cite{tsin00,colgin09}. The minimum which is exhibited
between the two peaks is to be associated to noise--induced firings that are
uncorrelated with the signal. The occurrence of an extra peak at low
frequency, which is also suggested by experimental data in \cite{N3}, is most
interesting, e.g., as a way to efficiently detect incoming signals along two
well defined noise levels. This seems to occur in nature and could also be
implemented in man--made devices.%
\begin{figure}
[tbh]
\begin{center}
\includegraphics[
height=6.6313cm,
width=12.0666cm
]%
{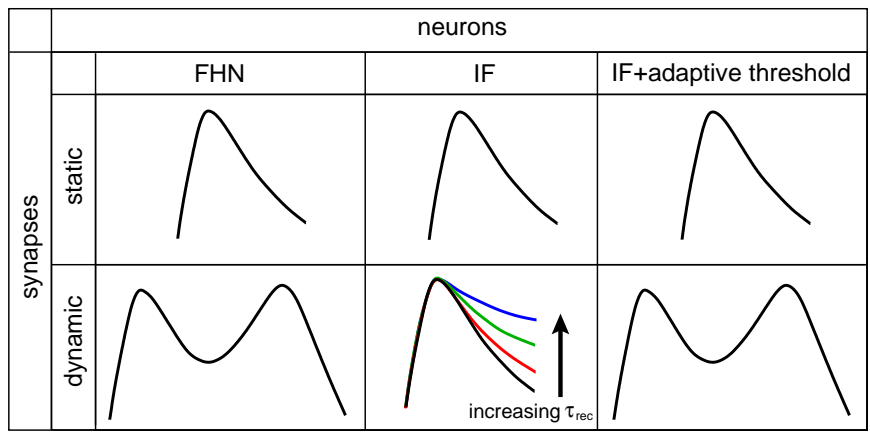}%
\caption{This shows schematically how the form of the information transfer
depends on assumptions concerning the neuron units and the synaptic links.
That is, an integrate--and--fire (IF) model unit can just produce a single
resonance, and the same is true for FitzHugh-Nagumo units and for IF units
with varying activation threshold as far as the connections are frozen.
Nevertheless, these two model units can produce two resonance peaks if the
connections are dynamic, e.g., if they show short--time fatigue plasticity
which is known to occur in many networks in nature.}%
\label{figure2}%
\end{center}
\end{figure}
The number of peaks and the frequency range at which they are located can
easily be controlled in the model by tuning parameter values, particularly
those concerning \textit{synaptic} dynamics.

Finally, we remark the model indication of two main ingredients of SMR. On one
hand, the system is expected to have activity--dependent excitability. This
may require short--term variations of intensity links in a networked system,
which is very common in practice \cite{NOSexcit2}. On the other hand, the
units in our model are able to adapt activation or firing thresholds to the
level of mean input. It is sensible to expect such adaptive thresholds
\cite{HH,mt}, and they have been observed recently in actual cortical regions
\cite{azouz}, for instance. A main conclusion is therefore that SMR should be
observed rather generally in neural media and in other excitable systems. We
summarize in figure \ref{figure2} the conditions in which such an interesting
phenomenon may occur. Incidentally, it is also worthwhile mentioning that the
present work adds to previous efforts analyzing the consequences in many
branches of science of the interplay between nonlinearities, signal and
forces, and environmental noise \cite{new}.

Supported by projects Junta de Andalucia FQM--01505 and MICINN--FEDER FIS2009--08451.


\begin{thebibliography}{99}
\bibitem {note0}We checked that allowing for some temporal correlations in the
signal does not modify essentially the phenomenon of interest here.
\bibitem {nota29bis}In practice, $M$ varies widely, namely, $M\in\left[
5,50\right]  $ depending on the neuron type and functionality; see, for
instance \cite{MA97,N4}. Well within the reported physiological range are also
our values for the link strength $J$ \cite{gil99} and for $u$ and $\tau_{rec}$
\cite{tsodyks97}.
\bibitem {nota1}This choice does not affect our conclusions. In fact, the same
behavior results in figure \ref{figure1} when using Poissonian signals as in
the experiment of interest \cite{yasuda08}.
\bibitem {nota2}Experimental data points in the figure have been transformed
using (\ref{nlr}) with $f_{0}=1$Hz, $\alpha=1.8$ Hz/dB, $A_{0}=73$ dB and
$\sigma_{A}=9$ dB. This choice does not affect essentially our findings,
however. E.g., using a different linear relationship at low and high $f$ (as
in \cite{yasuda08}) produces data that are well within the indicated error
bars. On the other hand, the Poissonian distributed pulses used for the empty
circles are of amplitude $\delta=0.4$ and mean frequency $f_{s}=3Hz$, and we
checked in a temporal window of $\Delta t=10ms$ after each pulse for a neuron
response to it, in order to maintain the closest similarity possible with the
experiment in \cite{yasuda08}.
\end{thebibliography}
\end{document}